\documentclass[conference]{IEEEtran}

\usepackage{flushend}
\usepackage{balance}
\usepackage{dblfloatfix}
\usepackage{flushend}
\usepackage{multicol}
\usepackage{algorithmic}
\usepackage{graphicx}
\usepackage{textcomp}
\usepackage{xcolor}
\usepackage{tcolorbox} 
\usepackage{caption}
\usepackage{subcaption}
\usepackage{booktabs} 
\usepackage{xurl}
\usepackage{xspace}
\usepackage{hyperref}
\usepackage{listings}
\usepackage[newfloat]{minted}
\usepackage{multirow}

\hypersetup{
    colorlinks=true,
    citecolor=black,
    linkcolor=black,
    filecolor=black,      
    urlcolor=black,
}
\definecolor{codegreen}{rgb}{0,0.6,0}
\definecolor{codegray}{rgb}{0.5,0.5,0.5}
\definecolor{codepurple}{rgb}{0.58,0,0.82}
\definecolor{backcolour}{rgb}{0.95,0.95,0.92}
\lstdefinestyle{mystyle}{
    backgroundcolor=\color{backcolour},   
    commentstyle=\color{codegreen},
    keywordstyle=\color{magenta},
    numberstyle=\tiny\color{codegray},
    stringstyle=\color{codepurple},
    basicstyle=\ttfamily\footnotesize,
    breakatwhitespace=false,         
    breaklines=true,                 
    captionpos=b,                    
    keepspaces=true,                 
    numbers=left,                    
    numbersep=5pt,                  
    showspaces=false,                
    showstringspaces=false,
    showtabs=false,                  
    tabsize=2
}
\newtcolorbox{boxH}{
    colback = white!90!gray, 
    colframe = black, 
    boxrule = 0pt, 
    leftrule = 3pt
}
\newcommand{\summary}[1]{ \begin{boxH} #1 \end{boxH} }

\newcommand{\eg}{\textit{e.g.,}~}
\newcommand{\ie}{\textit{i.e.,}~}

\AtBeginDocument{%
  \providecommand\BibTeX{{%
    \normalfont B\kern-0.5em{\scshape i\kern-0.25em b}\kern-0.8em\TeX}}}

\begin{document}

\title{PSASpotter: A Tool to Detect the Usage of Platform-Specific APIs in Python}

\author{
    \IEEEauthorblockN{Ricardo Job}
    \IEEEauthorblockA{
        \textit{UNINFO, IFPB}\\
        Cajazeiras, Brazil \\
        ricardo.job@ifpb.edu.br \\
    }
    \and
    \IEEEauthorblockN{Andre Hora}
    \IEEEauthorblockA{
        \textit{DCC, UFMG}\\
        Belo Horizonte, Brazil \\
        andrehora@dcc.ufmg.br \\
    }
}

\maketitle

\begin{abstract}
A \emph{platform-specific API} is implemented for a particular platform (\eg~operating system), thus, it may not work on other platforms than the target one.
Detecting the usage of such APIs is important for supporting software maintenance, as it allows maintainers to be alerted about APIs that could pose potential risks.
This paper proposes PSASpotter, a tool to detect the usage of platform-specific APIs in Python systems.
PSASpotter also identifies whether the platform-specific APIs are used within a defensive code, such as \texttt{try/except} blocks or \texttt{if} blocks that check the current platform.
PSASpotter can support the development of novel empirical studies about the usage of platform-specific APIs in the Python ecosystem.
Moreover, the defensive code detected by PSASpotter may contain alternative solutions for unavailable APIs, which can provide insights for software development and testing across multiple platforms.
PSASpotter is available at: \url{https://github.com/ricardojob/PSASpotter}.
Tool video: \url{https://youtu.be/d3WyozTAKS8}.
\end{abstract}

\maketitle

\section{Introduction}

A \emph{platform-specific API} is an API implemented for a particular platform (\eg~operating system) and therefore may not work on other platforms than the target one~\cite{job2024platform}.
Such limitations are typically associated with \emph{availability restrictions}, which, in this work, refer to conditions that limit the execution or availability of an API according to the underlying platform.
In Python, many APIs provided by the Python Standard Library have availability restrictions~\cite{job2024platform}.
For example, the API \texttt{os.pathconf}\footnote{\url{https://docs.python.org/3/library/os.html\#os.pathconf}} is available for Unix, while the API \texttt{\_thread.stack\_size}\footnote{\url{https://docs.python.org/3/library/\_thread.html\#thread.stack\_size}} is available for Windows.

\begin{figure}[h]
     \begin{subfigure}[b]{0.47\textwidth}
        \fbox{\includegraphics[width=\linewidth]{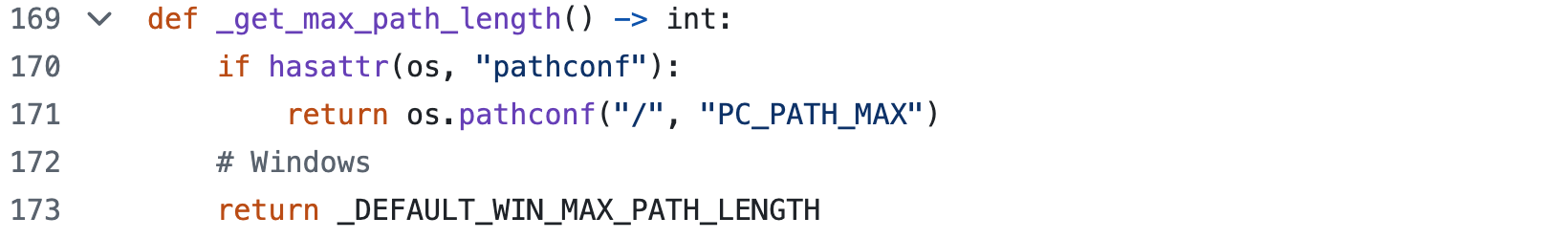}}
        \caption{Unix-specific API \texttt{os.pathconf}.}
        \label{fig:ray--l171}
     \end{subfigure}
    \hfill
     \begin{subfigure}[b]{0.47\textwidth}
        \fbox{\includegraphics[width=\linewidth]{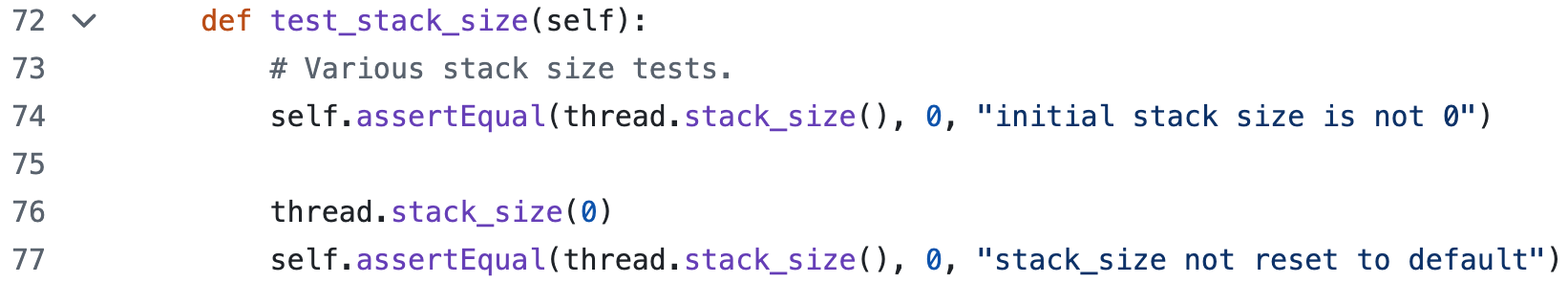}}
        \caption{Windows-specific API \texttt{\_thread.stack\_size}.}
        \label{fig:graalpython-l74}
     \end{subfigure}
     \caption{Examples of the platform-specific APIs in Python.}
     \label{fig:occurrences}
\end{figure}

A system that calls platform-specific APIs may implement defensive code to ensure it works properly on the desired platforms, such as \texttt{try/except} blocks or \texttt{if} blocks that check the API availability.
For instance, Figure~\ref{fig:ray--l171} presents an example where the Unix API \texttt{os.pathconf} is called within an \texttt{if} block that checks whether the current OS has such API.\footnote{\url{https://github.com/ray-project/ray/blob/fc98a5f286877ce7f6241961aca0c9127bee21ad/python/ray/tune/experiment/trial.py\#L169}}
Figure~\ref{fig:graalpython-l74} shows another case in which the Windows API \texttt{\_thread.stack\_size} is called, but the current platform is not checked.\footnote{\url{https://github.com/oracle/graalpython/blob/72c38809e2a4b1b8aeab475217fd4fb6b39ccea3/graalpython/lib-python/3/test/test_thread.py\#L72}}
As a result, this code might not work correctly when executed on Unix-like platforms.

Detecting the usage of platform-specific APIs is important for supporting software maintenance, as it allows project maintainers to be alerted about APIs that could pose potential risks.
However, this is not a trivial task because it requires 
(i) a comprehensive understanding of all platform-specific APIs and 
(ii) a tool to automatically identify them in source code.
In a prior study~\cite{job2024platform}, we found that the Python Standard Library~\cite{python_lib} has over 1,800 platform-specific APIs spread across 17 platforms, such as 
Linux, Windows, macOS, Unix, and Android.
We also found that platform-specific APIs are widely used in the Python ecosystem, with over 19K usages of 683 platform-specific APIs in the 100 analyzed projects~\cite{job2024platform}.

To overcome this gap, this paper proposes PSASpotter,\footnote{PSASpotter is available at: \url{https://github.com/ricardojob/PSASpotter}.} a tool to automatically detect the usage of platform-specific APIs in Python systems (Section~\ref{sec:tool}).
PSASpotter also identifies whether the platform-specific APIs are used within defensive code, such as \texttt{try/except} blocks or \texttt{if} blocks that check the current platform.
Next, we discuss practical applications of tool (Section~\ref{sec:applications}).
Finally, we evaluate the tool's performance in detecting whether platform-specific API usage occurs within defensive code (Section~\ref{sec:study}).
The tool achieves precision, recall, and accuracy of over 97\%, 83\%, and 87\%, respectively.

PSASpotter can support the development of novel empirical studies about the usage of platform-specific APIs.
To support such studies, we mined the usage of platform-specific APIs in 9,205 Python repositories and created a large-scale dataset.
Additionally, the defensive code detected by PSASpotter may contain alternative solutions for unavailable APIs.
These alternative solutions can provide insights for software development and testing across multiple platforms.

\noindent\textbf{Novelty:}
To our knowledge, this is the first tool to detect the usage of platform-specific APIs. 
We also provide three practical applications
and evaluate its performance in identifying platform-specific API usage within defensive code.

\section{PSASpotter}
\label{sec:tool}

\subsection{Overview}


PSASpotter is a tool to automatically detect the usage of platform-specific APIs in Python systems.
Specifically, it identifies 1,841 platform-specific APIs from the Python Standard Library~\cite{python_lib}.
These APIs were derived from our previous research, where we mined the entire Python documentation to identify platform-specific APIs~\cite{job2024platform}.
The tool can be used via the command line, receiving the system to be analyzed as input and providing the API usage occurrences as output.

\subsection{Detecting Platform-Specific API Usage}
\label{sec:extracting}

\begin{figure}[h]
     \centering
     \fbox{\includegraphics[width=\linewidth]{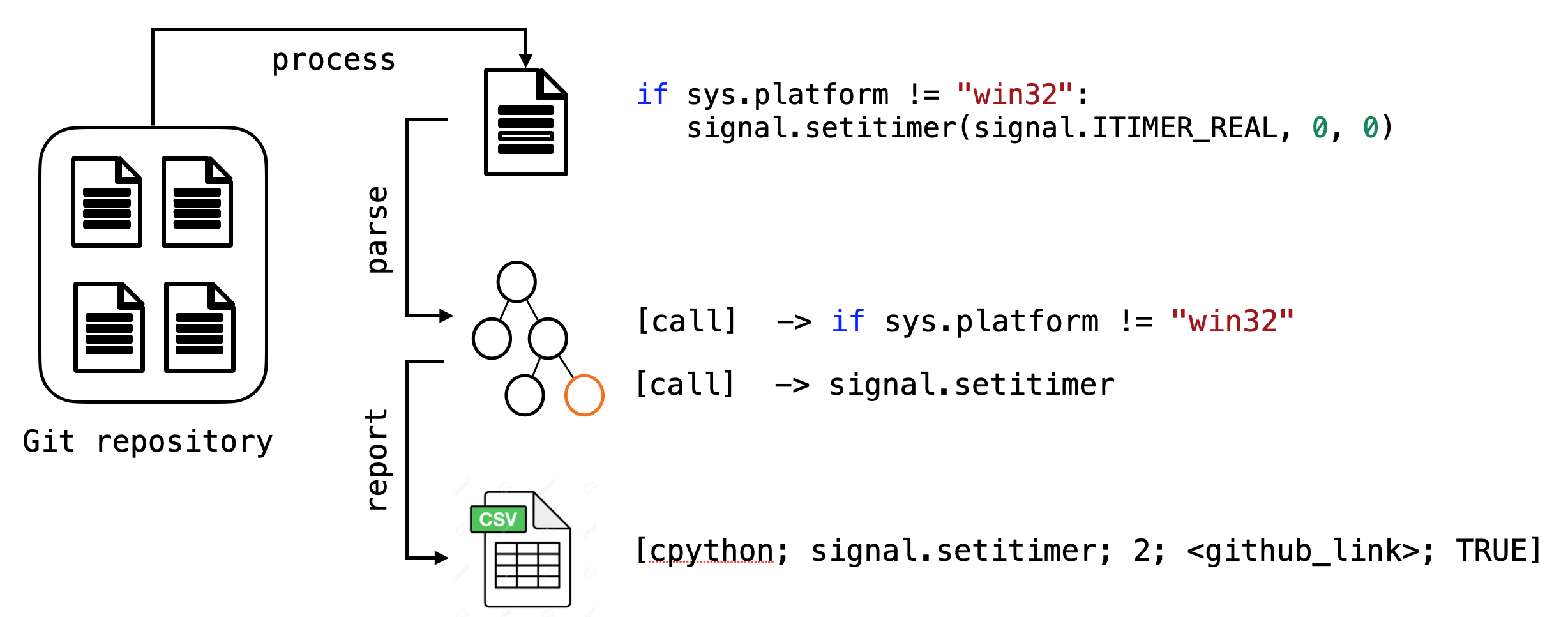}}
         \caption{Overview workflow of the PSASpotter.}
        \label{fig:psaspotter-overview}
\end{figure}

PSASpotter is an AST-based tool that detects API usage at the function/method level.
Given a Git project, PSASpotter analyzes all Python files, converts them into an AST (Abstract Syntax Tree), identifies calls to platform-specific APIs, and exports details of their usage. 
Specifically, for each usage, PSASpotter reports information about the analyzed project (name and commit), the used API (name and availability), and the usage location (filename, line, and GitHub link\footnote{Example of GitHub link: \url{https://github.com/ray-project/ray/blob/fc98a5f286877ce7f6/python/ray/_private/ray_process_reaper.py\#L30}}).
In addition, PSASpotter reports whether the usage occurs within defensive code.
Figure~\ref{fig:psaspotter-overview} provides an overview of the tool for detecting platform-specific API usages.

\subsection{Detecting Usage within Defensive Code}
\label{sec:calculate}

\begin{figure}[h]
     \begin{subfigure}[b]{0.47\textwidth}
        \fbox{\includegraphics[width=\linewidth]{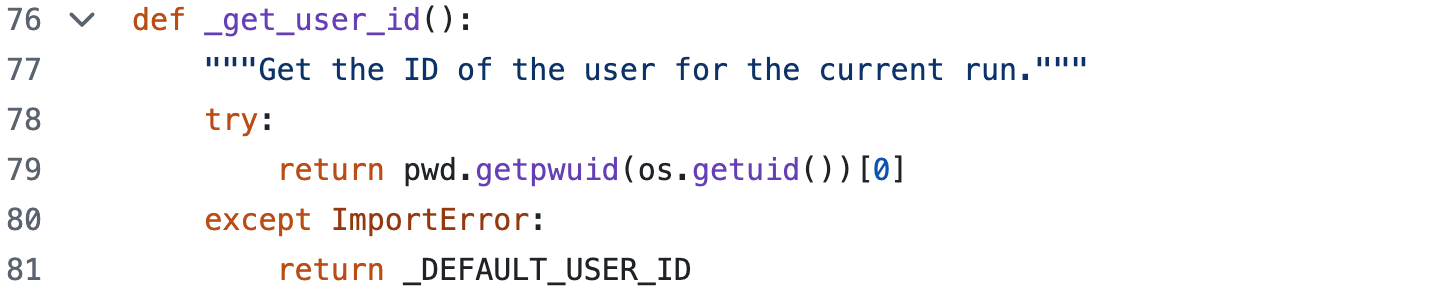}}
        \caption{Platform-specific API \texttt{os.getuid} used within a \texttt{try/\-except} block.}
        \label{fig:mlflow-l79}
     \end{subfigure}
     \begin{subfigure}[b]{0.47\textwidth}
        \fbox{\includegraphics[width=\linewidth]{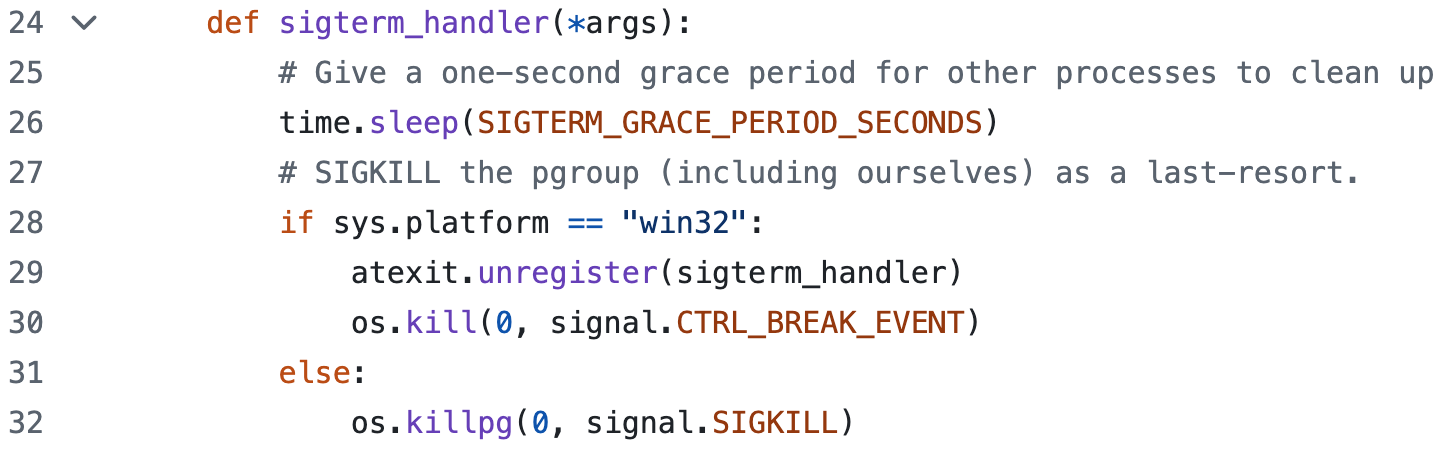}}
        \caption{Platform-specific API \texttt{os.kill} used within an \texttt{if} block that checks the current platform. In this case, \texttt{sys.platform} is used to verify if the OS is Windows.}
        \label{fig:ray-l32}
     \end{subfigure}
     \begin{subfigure}[b]{0.47\textwidth}
        \fbox{\includegraphics[width=\linewidth]{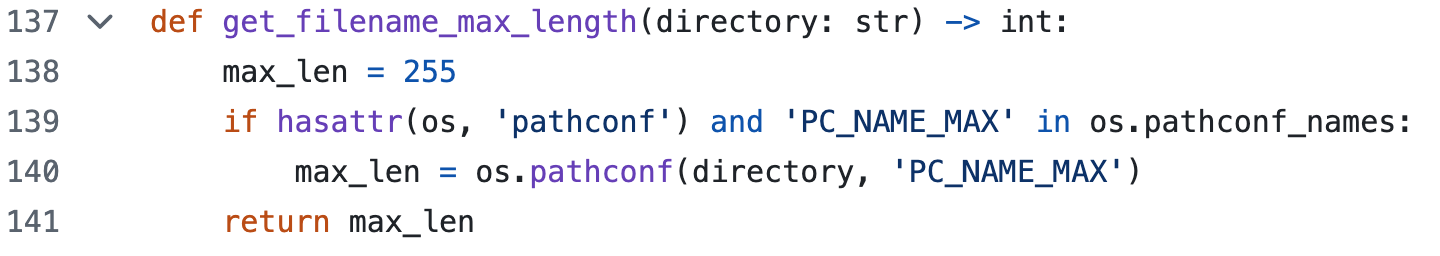}}
        \caption{Platform-specific API \texttt{os.pathconf} used within an 
        \texttt{if} block that checks the existence of the API in the current platform. In this case, \texttt{hasattr} is used to verify if \texttt{os} has the attribute \texttt{pathconf}.}
        \label{fig:ray-l171}
     \end{subfigure}
     \caption{Platform-specific APIs usage within defensive code.}
     \label{fig:detecting}
\end{figure}

PSASpotter also identifies whether platform-specific APIs appear within defensive code, checking four cases:

\begin{itemize}
    
    \item \textbf{\texttt{try/except} blocks:} Platform-specific APIs are called within \texttt{try/except} blocks. This happens when developers want to avoid any exceptions caused by calling the API on the wrong platform (\eg Figure~\ref{fig:mlflow-l79}).

    \item \textbf{\texttt{if} blocks that check current platform:} Platform-specific APIs are called within \texttt{if} blocks that identify the current platform (such as Unix or Windows). In this case, we consider all 15 APIs available in the Python Standard Library to identify the current platform~\cite{jobandhora2025}, such as \texttt{sys.platform}\footnote{\url{https://docs.python.org/3/library/sys.html\#sys.platform}} and \texttt{os.name}\footnote{\url{https://docs.python.org/3/library/os.html\#os.name}} (\eg Figure~\ref{fig:ray-l32}).

    \item \textbf{\texttt{if} blocks that verify API existence:} Platform-specific APIs are called within \texttt{if} blocks that check the existence of the API in the current platform. In this case, we consider the built-in function \texttt{hasattr}\footnote{\url{https://docs.python.org/3/library/functions.html\#hasattr}} (\eg Figure~\ref{fig:ray-l171}).

    \item \textbf{\texttt{@skipif} decorators in tests:} When the usage of platform-specific APIs is located in tests, PSASpotter verifies whether these tests are annotated with decorators to conditionally skip (\eg~\texttt{@skipif}) their execution under specific conditions.
    Using such decorators also represents a form of defensive programming, as they allow tests to be conditionally skipped on platforms where they are not supported.
    The tool supports five decorators, two in pytest and three in unittest: \texttt{@pytest.mark.\-skipif}, \texttt{@pytest.\-mark.xfail}, \texttt{@unittest.\-skipIf}, \texttt{@unittest.\-skip\-Unless}, and \texttt{@unit\-test.\-expected\-Failure}.
    
\end{itemize}


\subsection{Tool Usage}

PSASpotter can be used via the command line.
First, we install it via pip (\texttt{pip install psaspotter}).
Next, we can use the tool to extract information about the usage of platform-specific APIs in a Git repository:

\begin{lstlisting}[language=bash]
# Running PSASpotter
$ psaspotter <git_repo> -o <output_file.csv>
\end{lstlisting}

The first parameter is the Git repository to be analyzed.
The parameter \texttt{-o} defines the name of the CSV output file, 
while the parameter \texttt{-c} specifies a commit or tag. 
For example, we can analyze a specific commit or tag with the parameter \texttt{-c}.
PSASpotter also supports filtering APIs by category of platform-specific APIs using the parameter \texttt{-p}, such as main OSs (Linux, Windows, macOS, and Unix) and proprietary OSs (Solaris, AIX, and VxWorks).
To improve maintainability, the tool uses a predefined list of platform-specific APIs from previous research by default, 
but it also supports the parameter \texttt{-f}, which allows users to provide a JSON file with the list of platform-specific APIs to be detected.

\section{Practical Applications}
\label{sec:applications}

\subsection{Novel Empirical Studies and Datasets}

We foresee that PSASpotter can support novel empirical studies about the usage of platform-specific APIs.
For instance, it can reveal the diversity of platform-specific API usage and its potential risk when not used with proper defensive code.

In our previous research~\cite{job2024platform}, we proposed an empirical study on the availability and usage of platform-specific APIs in 100 Python projects. 
Leveraging PSASpotter, we substantially expanded this dataset to 9,205 Python repositories. 
Table~\ref{tab:overview} shows some statistics about the new dataset, which contains 709,191 usages of platform-specific APIs, including 465,848 (66\%) in production files and 243,343 (34\%) in test files.



\begin{table}[h]
\centering
\caption{Dataset statistics.}
\label{tab:overview}
\begin{tabular}{ll}
\toprule
\textbf{Data} & \textbf{Value}  \\
\midrule
GitHub repositories & 9,205 \\
\midrule
Usage Platform-Specific APIs: all files & 709,191 \\
Usage Platform-Specific APIs: production files & 465,848 \\
Usage Platform-Specific APIs: test files & 243,343 \\
\midrule
Processed Files: all files & 1,953,333 \\
Processed Files: production files & 1,449,103 \\
Processed Files: test files & 504,230 \\
\bottomrule
\end{tabular}
\end{table}


\summary{
\textbf{Application 1:}
PSASpotter can support the development of novel empirical studies about the usage of platform-specific APIs in Python.
To support such studies, we mined the usage of platform-specific APIs in 9,205 Python repositories and created a large-scale dataset, which is publicly available at: \url{https://doi.org/10.5281/zenodo.17857593}.}

\subsection{Detect Alternative Platform-Specific APIs}

\begin{figure}[h]
     \centering
         \fbox{\includegraphics[width=0.45\textwidth]{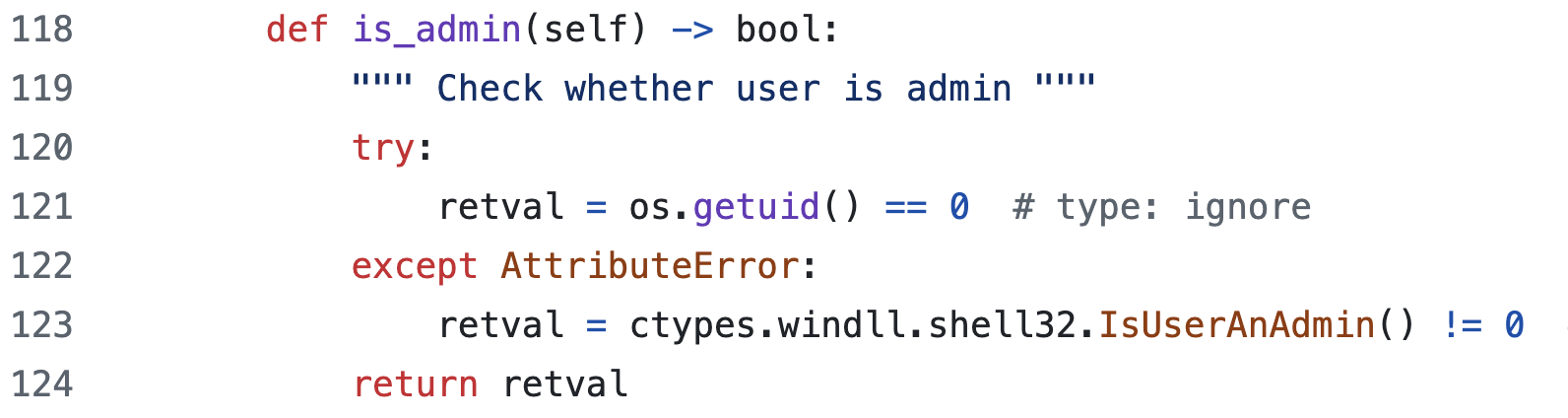}}
         \caption{Alternative to \texttt{os.getuid} (FaceSwap).}
        \label{fig:is_admin}
\end{figure}

When calling platform-specific APIs within defensive code, developers may provide an alternative solution in case the API is unavailable.
For example, they may call a third-party API or implement in-house solutions to handle the missing API.
These alternative solutions can serve as a valuable resource for identifying and addressing potential challenges when developing and testing across multiple platforms.
For instance, in project FaceSwap, the method \texttt{is\_\-admin}\footnote{\url{https://github.com/deepfakes/faceswap/blob/cbaad146d5aca9bd714bb6c69b10dd7c02d88f9d/setup.py\#L121}} calls the Unix-specific API \texttt{os.getuid}\footnote{\url{https://docs.python.org/3/library/os.html\#os.getuid}} in a \texttt{try} block, as presented in Figure~\ref{fig:is_admin}.
In case an exception is raised, the alternative Windows API \texttt{ctypes.\-windll.\-shell32.\-IsUserAnAdmin} is used instead.\footnote{\url{https://docs.python.org/3.11/library/ctypes.html}}
Finally, Figure~\ref{fig:ray-l32} presents another potential alternatives: \texttt{os.kill} and \texttt{os.killpg}.

\summary{
\textbf{Application 2:}
The defensive code detected by PSASpotter may contain alternative solutions for unavailable APIs.
These alternative solutions can provide insights for identifying and resolving potential challenges encountered during development and testing across multiple platforms.
}

\subsection{Identify Libraries and Frameworks with Dependence on Platform-Specific APIs}

Detecting the usage of platform-specific APIs may be particularly important for libraries and frameworks.
These projects are typically executed and tested on multiple platforms by their users.
Therefore, having defensive code when using platform-specific APIs is fundamental to reducing the chances that errors propagate to users.
As an example, Table~\ref{tab:dependents} presents five highly relevant\footnote{\url{https://lp.jetbrains.com/python-developers-survey-2023/}} libraries and frameworks in the Python ecosystem with their number of dependent repositories in GitHub and usage of platform-specific APIs.

\begin{table}[h]
\centering
\small
\caption{Usage of platform-specific APIs in selected Python projects.}
\label{tab:dependents}
\begin{tabular}{lrr}
\toprule
\multirow{2}{*}{\textbf{Project}} & \textbf{Dependent} & \textbf{Usage of Platform-} \\ 
& \textbf{Repositories} & \textbf{Specific APIs } \\
\midrule
django/django &   1,732,014 &305\\
psf/requests &   3,086,331 &96 \\
fastapi/fastapi &   436,052 &24 \\
pallets/flask &   2,075,961 &10 \\
encode/django-rest-framework &   710,636 & 4\\
\bottomrule
\end{tabular}
\end{table}

\summary{
\textbf{Application 3:}
PSASpotter can identify the usage of platform-specific APIs in libraries and frameworks, helping developers decide whether such calls should rely on defensive programming and potentially reducing the chances that errors propagate to users.}


\section{Tool Performance}
\label{sec:study}

PSASpotter identifies whether these APIs are used outside defensive code structures. 
To evaluate this capability, we assess the performance of the tool in identifying whether platform-specific API usage occurs within defensive code.

\subsection{Case Study}
\label{sec:selecting-systems}
We selected 100 popular Python real-world software systems projects hosted on GitHub.
We relied on our prior dataset to collect software projects~\cite{job2024platform, platform-api-zenodo}.
We chose this dataset of projects due to the following characteristics: 
(1) they are written in Python, 
(2) they are real-world software systems, and 
(3) they use platform-specific APIs.
This dataset includes widely used projects, such as Numpy, Django, and Pandas.

\subsection{Manual Flagging}
\label{sec:flagging}

Initially, we collected 1,544 occurrences used by the selected client systems.
Among these occurrences, 977 were located in the production code, while 567 happened in the test code.
Next, for each file where an occurrence appeared, we manually examined and classified based on the usage context of platform-specific API.
This process resulted in 626 instances identified as occurring in defensive code, while 918 occcurred without defensive code.
We analyzed whether each file contained protective handling around the function where the API call occurs.
This evaluation is publicly available at: \url{https://doi.org/10.5281/zenodo.14018574} 

\subsection{Evaluation: Precision, Recall, and Accuracy}



Table~\ref{tab:performance} details the precision, recall, and accuracy of the 1,544 occurrences.
It also shows metrics for the top-5 projects with the most uses.
We manually flagged 626 occurrences as defensive code and 918 as the absence of defensive code.
After running the tool, we observed a large number of false negatives (171) and few false positives (23). 
Regarding the correct predictions, we observed 895 true positives and 455 true negatives.
Overall, we find a precision of 97.49\%, a recall of 83.96\%, and an accuracy of 87.44\% for defensive code identification. 
The overall precision close to 98\% indicates that PSASpotter makes correct predictions in nearly all cases.
Note that the overall recall is not low (83.96\%), indicating that it often correctly identify defensive code.
Finally, the accuracy at 87.44\% shows that the tool provides good, reliable predictions.

\begin{table}[h]
\centering
\caption{Performance of defensive code identification.}
\label{tab:performance}
\begin{tabular}{lccccc}
\toprule
 \textbf{Project} &  \textbf{Correct}& \textbf{Incorrect}& \textbf{Prec.} & \textbf{Rec.} & \textbf{Acc.} \\
\midrule
 All  &  1,350& 194& 97.49& 83.96& 87.44\\
\midrule
 Test        &  491 & 76  & 97.07 & 84.88 & 86.60 \\
 Production  &  859 & 118 & 97.79 & 83.33 & 87.92 \\
\midrule
 saltstack/salt  &  214 & 75 & 99.21 & 63.00 & 74.05 \\
 ansible/ansible  &  166 & 7 & 97.65 & 94.32 & 95.95 \\
 ray-project/ray  &  107 & 11 & 91.53 & 90.00 & 90.68 \\
 kovidgoyal/kitty  &  64 & 2 & 98.04 & 98.04 & 96.97 \\
 spotify/luigi  &  51 & 6 & 92.00 & 95.83 & 89.47 \\
\bottomrule
\end{tabular}
\footnotesize{Precision (Prec.), Recall (Rec.), and  Accuracy (Acc.)}\\
\end{table}

\summary{ \textbf{Results}:
PSASpotter has a precision of 97.49\%, recall of 83.96\%, and accuracy of 87.44\% 
to identifying whether the platform-specific API usage occurs within defensive code.}
\subsection{Main Reasons of Divergence}
\label{sec:discussion}

Our findings show that PSASpotter performs well in detecting whether platform-specific API usage occurs within defensive code. 
To understand the discrepancies between manual inspection and the tool’s output, we examined the 194 conflicting cases and identified three main reasons for divergence (another 26 cases are related to manual flagging).

\noindent \textbf{In-house solutions}. 
In some cases, projects employ in-house solutions to identify the operating system.
We identified 34 involving internal decorator and 47 occurrences involving internal APIs, implemented specifically for each project (\ie 
\texttt{@t.skip.if\_win32}\footnote{\url{https://github.com/celery/celery/blob/fe762c3a26e56ff34608244fc04336b438f8fa0c/t/unit/apps/test_multi.py\#L389}} 
and 
\texttt{salt.utils.platform.is\_windows}\footnote{\url{https://github.com/saltstack/salt/blob/53db41263216ba00a94e10fa2fceefa732f2b776/salt/master.py\#L1241}}).

\noindent \textbf{Distinct values due to platform-specific APIs}. 
We also found 21 occurrences involving value assignments, 17 involving value access, and 18 related to handling within the caller function. 
This reason is further supported by existing literature~\cite{jobandhora2025}.
These cases involve control flow data that would require dynamic analysis for accurate detection.

\noindent \textbf{Guard clauses}.
We also found 31 occurrences related to the guard clause.
The tool was unable to link the API calls to the guard clause because the calls did not occur within the same \texttt{if} block.
Notice that these cases are very rare.

\section{Limitations}

PSASpotter works for Python systems, and the platform-specific APIs come from the Python Standard Library~\cite{python_lib}.
Such library, which describes the standard library distributed with Python, is very extensive and offers a wide range of facilities~\cite{python_lib}.
Thus, it is used by virtually every Python application.
Despite that, other sources of platform-specific APIs may exist in the Python ecosystem.
Further studies could investigate other ways to identify the current platform, for example, from popular libraries and frameworks.
Therefore, PSASpotter can be extended by incorporating more platform-specific APIs from additional sources and improving accuracy in identifying usage within defensive code.
Moreover, further tools can be built to explore platform-specific APIs in other programming languages and analyze dynamic API calls.


\section{Related Work}

Application Programming Interface (API) is a widely studied research topic in software engineering~\cite{principles-design-shared-software, code-samples-fram, effect-oop-framework-productivity,  measuring-lib-stability, robillard_automated_2013}. 
Prior research has empirically examined API evolution, breaking changes, deprecation, and migration~\cite{sawant_reaction_2018, emse2020_aline, sqj2017,  kula2018developers,  linares2013api, mahmud2022android2, nascimento_exploring_2022, xavier2017historical, xia2020android, barbosa2022and, martinez2020and}.
Several tools have also been proposed to support API maintenance and evolution, including SemDiff~\cite{dagenais2009semdiff}, apiwave~\cite{hora2015apiwave}, APIDiff~\cite{brito2018apidiff}, Aexpy~\cite{du2022aexpy}, REPFINDER~\cite{huang2021repfinder}, and Deprewriter~\cite{ducasse2022deprewriter}, to name a few.
However, none of these studies or tools address the detection or analysis of platform-specific API usage, leaving the frequency, scope, and risks associated with this phenomenon unexplored.
To address this gap, PSASpotter can support investigations into the extent to which platform-specific APIs are handled and used.

In the context of platform-specific APIs, we have empirically explored their availability and usage in client systems~\cite{job2024platform}.
Recently, we investigated OS-specific tests, namely, a test that identifies the operating system on which they are executed~\cite{jobandhora2025, job2025os-test-cpython}.
However, neither line of research provides automated support for detecting code dealing with platforms.
PSASpotter contributes to the API literature by providing an automatic tool for detecting the usage of platform-specific APIs and supporting novel research on this topic.

\section{Conclusion}

This paper proposed PSASpotter, a tool to detect the usage of platform-specific APIs in Python systems.
PSASpotter also identifies whether the platform-specific APIs are used within defensive code.
In addition, we provided three practical applications for the usage of this tool.
Moreover, we evaluated the performance of the tool in identifying whether the platform-specific API usage occurs within defensive code.
Lastly, we discussed practical applications related to development and testing across multiple platforms.

\section*{Acknowledgments}

This research is supported by CAPES, CNPq, and FAPEMIG.

\bibliographystyle{plain}
\bibliography{main}
\end{document}